\documentclass{andp2012}
\usepackage[english]{babel}
\title{Loschmidt echo and fidelity decay near an exceptional point}
\author[S: Longhi]{Stefano Longhi \inst{1,}\footnote{Corresponding author\quad E-mail:~\textsf{longhi@fisi.polimi.it}}}
\address[1]{Dipartimento di Fisica, Politecnico di Milano and Istituto di Fotonica e Nanotecnologie del Consiglio Nazionale delle Ricerche, Piazza L. da Vinci 32, I-20133 Milano, Italy}
\shortauthors{S. Longhi }
\begin{abstract}
Non-Hermitian classical and  open quantum systems near an exceptional point (EP) are known to undergo strong deviations in their dynamical behavior under small perturbations or slow cycling of parameters as compared to Hermitian systems. Such a strong sensitivity is at the heart of many interesting phenomena and applications, such as the asymmetric breakdown of the adiabatic theorem, enhanced sensing, non-Hermitian dynamical quantum phase transitions and photonic catastrophe. Like for Hermitian systems, the sensitivity to perturbations on the dynamical evolution can be captured by Loschmidt echo and fidelity after imperfect time reversal or quench dynamics. Here we disclose a rather counterintuitive phenomenon in certain non-Hermitian systems near an EP, namely the deceleration (rather than acceleration) of the fidelity decay and improved Loschmidt echo as compared to their Hermitian counterparts, despite large (non-perturbative) deformation of the energy spectrum introduced by the perturbations. This behavior is illustrated by considering the fidelity decay and Loschmidt echo for the single-particle hopping dynamics on a tight-binding lattice under an imaginary gauge field. 
\end{abstract}
\shortabstract
\begin{document}
\maketitle

\section{Introduction}
Exceptional points (EPs) are special spectral degeneracies
of non-Hermitian Hamiltonians governing
the dynamics of open classical and quantum systems \cite{R1,R2,R3,R4,R5}.
Recently, the dynamical behavior of non-Hermitian systems near an EP has sparked a great interest with a wealth of
applications in several areas of physics, notably in integrated photonics systems \cite{R6,R7,R8,R9,R10,R11,R12,R13}, acoustics \cite{R14,R15,R16} and optomechanics \cite{R17,R18,R19,R20,R21,R22} to mention a few (for recent reviews and more extended references see \cite{R23,R24,R25,R26,R26bis,R26tris}).
At the EP, two or more eigenvalues and the corresponding
eigenstates of the Hamiltonian $H$ coalesce. An ubiquitous property of a non-Hermitian system is its extreme sensitivity to perturbations when operated closed to an EP.
Such a result stems from the fact that, if the Hamiltonian $H=H(h)$ depending on a control parameter $h$ shows an EP at $h=h_0$, then the energy spectrum $E=E(h)$ and corresponding eigenfunctions are non-analytic and show a branch point at $h_0$, with $(d E/ d h)_{h_0}=\infty$ \cite{R1,R27,R28}. The strong sensitivity to perturbations is at the heart of  several phenomena studied in recent works, such as sensing enhancement in optical micro cavities \cite{R11,R12,R29,R30,R31}, cavity-assisted enhanced metrology and quantum sensing \cite{R32,R32bis}, ultra sensitive micro-scale gyroscopes \cite{R33,R34,R35}, quantum and photonic catastrophes \cite{R36,R37}, critical phenomena and dynamical quantum phase transitions \cite{R38,R38bis,R38uff,R38tris}. A related phenomenon observed as the parameter $h$ is slowly cycled around an EP is the asymmetric breakdown of the adiabatic theorem and unidirectional transitions \cite{R18,R36,R37,R38,R39,R40,R41,R42,R43,R44,R45}, resulting in topological energy transport \cite{R18} and asymmetric mode switching \cite{R43}.\\
In the study of complex quantum systems, the stability of quantum evolution in the presence of perturbations is rather generally measured by quantities such as Loschmidt echo and fidelity after imperfect time reversal or quench dynamics \cite{R46,R47,R48,R48bis}. For example, in Loschmidt echo setups an initial
state is propagated for a given time and then reversed. The comparison of the resulting state and the initial one constitutes a measure of the irreversibility suffered by the system during its evolution and generated by differences between the forward and backward dynamics. Likewise, an initial state $\psi_0$ evolves, after a time interval $t$,  into the two states $\psi_1(t)$ and $\psi_2(t)$ under the Hamiltonians ${H}_1$ and ${H}_2={H}_1+ \epsilon P$, where $\epsilon P$ is a perturbation: the overlapping $\mathcal{F}(t) = |\langle \psi_2 (t) | \psi_1 (t) 
\rangle|^2$, referred to as fidelity,  provides a measure of the stability of the dynamics under the perturbation.  When we extend such concepts to non-Hermitian dynamical systems, the effect of a perturbation on the dynamical behavior is expected to be strongly enhanced near an EP, resulting in a degraded fidelity in short times. In this work it is shown that, rather counterintuitively, in certain perturbed non-Hermitian models the fidelity decay can be {\it decelerated} (rather than accelerated) as the system operates {\it closer} to (rather than far apart from) an EP. We illustrate such an intriguing behavior by considering a paradigmatic model of non-Hermitian transport in tight-binding lattices with asymmetric hopping, namely the Hatano-Nelson model \cite{R49,R50,R51}. This model shows a rich physics and has seen a renewed interest very recently \cite{R52,R53,R54,R55,R56,R57,R58,R59,R60,R61,R62,R63,R64,R65,R66,R66bis,R67,R68,R69,R70}.    

\section{Model and non-Hermitian stationary perturbation theory}
Let ${H}_0$ be a Hermitian $N \times N$ matrix that describes rather generally {the coherent hopping dynamics of a single-particle} on a finite-dimensional tight-binding network, and let us indicate by $E_1$, $E_2$,..., $E_N$ and ${\bf u}_1$, ${\bf u}_2$,..., ${\bf u}_N$ the eigen-energies and corresponding eigenvectors of ${H}_0$, i.e.
\begin{equation}
{H}_0  {\bf u}_n = E_n { \bf u}_n 
\end{equation}
($n=1,2,3,...,N$). For the sake of simplicity, we assume that the eigenvalues are not degenerate, take the normalization $\langle {\bf u}_m | {\bf u}_n \rangle = \delta_{n,m}$ for the eigenvectors, and 
assume short-range hopping so that $({H}_0)_{n,m}=0$ for $|n-m| >L$ for some integer $L \geq 1$. Indicating by ${X}$ the $N \times N$ non-unitary diagonal matrix given by 
\begin{equation}
{X}_{n,m}= \exp(-hn) \delta_{n,m}
\end{equation}
 with $h \geq 0$,
 we can introduce the {\it pseudo-Hermitian} Hamiltonian ${H}_1$ via the similarity transformation
\begin{equation}
{H}_1 ={X} {H}_0 {X}^{-1},
\end{equation}
i.e.
\begin{equation}
\left( {H}_1 \right)_{n,m}=\left( {H}_0 \right)_{n,m} \exp [ h(m-n) ].
\end{equation}
The similarity transformation basically corresponds to a non-Hermitian gauge transformation of the wave function, which arises by application of a synthetic imaginary gauge field $h$. Such an imaginary gauge phase could be realized experimentally in photonic microring structures and in ultracold atomic systems, as proposed in some recent works \cite{R53,R54,R64,R66}.
 For example, if ${H}_0$ describes the hopping dynamics in a uniform one-dimensional chain with nearest-neighbor hopping amplitude $\kappa$ and open boundary conditions, i.e.
 \begin{equation}
 {H}_0=\left(
 \begin{array}{ccccccc}
 0 & \kappa & 0 & 0 & ... & 0 & 0 \\
 \kappa & 0 & \kappa& 0 & ... & 0 & 0 \\
 0 & \kappa & 0 & \kappa & ... & 0 & 0 \\
 ... & ... & ... & ...& ...& ... & ... \\
  0 & 0 & 0 & 0 & ... & 0 & \kappa \\
 0 & 0 & 0 & 0 & ... & \kappa & 0 
 \end{array}
 \right)
 \end{equation}
 a non-vanishing imaginary gauge phase $h$ introduces asymmetric forward/backward hopping amplitudes $\kappa_1= \kappa \exp(h)$ and $\kappa_2= \kappa \exp(-h)$  in the pseudo-Hermitian Hamiltonian ${H}_1$, namely one has
  \begin{equation}
 {H}_1=\left(
 \begin{array}{cccccc}
 0 & \kappa \exp(h) & 0 &   ... & 0 & 0 \\
 \kappa \exp(-h)& 0 & \kappa \exp(h)&  ... & 0 & 0 \\
 0 & \kappa \exp(-h) & 0 & ... & 0 & 0 \\
 ... &  ... & ...& ...& ... & ... \\
  0 & 0 & 0 & ... & 0 & \kappa \exp(h) \\
 0 & 0 & 0 & ... & \kappa \exp(-h) & 0 
 \end{array}
 \right).
 \end{equation}
  This pseudo-Hermitian Hamiltonian reproduces the Hatano-Nelson model without disorder \cite{R49} and shows interesting topological properties, as shown in recent works \cite{R64}.  
Clearly, the Hamiltonians ${H}_0$ and  ${H}_1$ are isospectral, and the eigenvectors $\bf{v}_n$ of ${H}_1$ are simply given by ${\bf v}_n=X{ \bf{u}_n}$, i.e. ${H}_1 {\bf{v}}_n =E_n {\bf{v}}_n$ with
\begin{equation}
({\bf{v}}_n)_l=({\bf{u}}_n)_l \exp(-lh)
\end{equation}
($l,n=1,2,...,N$). Note that the imaginary gauge field squeezes the eigenstates closer to the left edge (for $h>0$), i.e. all eigenstates become {\em left-edge states}. This effect has been referred in some recent works to as the {\em non-Hermitian skin effect}.
Since ${H}_1$ is not Hermitian, the left and right eigenvectors of ${H}_1$ do not coincide. Indicating by $\bf{v}^{\dag}_n$ the eigenvector of the adjoint ${H}_1^{\dag}=X^{-1} {H}_0 X$ with energy $E_n$, one has
\begin{equation}
({\bf{v}^{\dag}}_n)_l=({\bf{u}}_n)_l \exp(lh).
\end{equation}
Note that the ratio
\begin{eqnarray}
\frac{\langle {\bf v}_n | {\bf v}_n \rangle \langle {\bf v}_n^{\dag} | {\bf v}_n^{\dag} \rangle}{| \langle {\bf v}^{\dag}_n | {\bf v}_n \rangle |^2} & = &  \left( \sum_{l=1}^{N} \left| ({\bf u}_n)_l \right|^2 \exp(2hl) \right) \nonumber \\
& \times & 
\left( \sum_{l=1}^{N} \left| ({\bf u}_n)_l \right|^2 \exp(-2hl) \right)
\end{eqnarray}
is one in the Hermitian limit $h=0$, while it increases like $\sim 1/ \alpha^{2N}$ and diverges as $h \rightarrow \infty$, where we have set $\alpha \equiv \exp(-h)$: this is the signature that an EP is approached as $h$ is increased. This can be also shown by direct computation of the matrix ${H}_{1}$ in the large $h$ limit. In this case, the dominant elements of ${H}_{1}$ are those on the diagonal ${m=n+L}$, which scale like $\sim \exp(hL)$ according to Eq.(4). Hence, in the large $h$ limit, at leading order in the small parameter $\alpha$ one has ${H}_1 = \exp(hL) \left[ {A}+ O(\alpha) \right]$ with $({A})_{n,m}= ({H}_0)_{n,m} \delta_{n,m-L}$. Clearly, since the matrix ${A}$ has an EP at zero energy of order $(N-L)$, in the large $h$ limit such an EP is approached by the matrix ${H}_1$ as well.\par
Let us now consider how a perturbation affects the spectrum of the pseudo-Hermitian Hamiltonian ${H}_1$. Let us consider the perturbed Hamiltonian
\begin{equation}
{H}_2={H}_1+ \epsilon P
\end{equation} 
where the elements of the matrix perturbation $P$ are of the same order than the elements of ${H}_0$, say of order $\sim 1$, while $\epsilon$ is a small parameter that measures the strength of the perturbation. Clearly, $H_2$ is isospectral to the matrix
\begin{equation}
H_0^{\prime}=X^{-1} H_2 X=H_0+\epsilon P^{\prime}
\end{equation}
where we have set
\begin{equation}
P^{\prime} \equiv X^{-1} P X.
\end{equation}
Hence we can compute the energy spectrum of $H_0^{\prime}$, which differs from the Hermitian Hamiltonian $H_0$ by the (generally non-Hermitian) perturbation term $\epsilon P^{\prime}$. If one applies standard Rayleigh-Schr\"odinger perturbation theory for non-degenerate eigenvalues, at first-order in $\epsilon$ the varied eigenvalues of $H_2$ are thus given by
\begin{equation}
E^{\prime}_n \simeq E_n+ \epsilon \langle {\bf u}_n| P^{\prime} | {\bf u}_n \rangle.
\end{equation}
Clearly, while Eq.(13) holds in the Hermitian limit $h=0$, it rapidly fails to predict the correction of the eigenvalues in the non-Hermitian case when the perturbation $P$ is long-range and the number $N$ of sites of the network is large enough. In fact, from Eq.(12) it readily follows that, for a long-range perturbation such as, for example,  the element $P_{N,1}$ does not vanish, one has $P^{\prime}_{N,1}=P_{N,1} \exp[h(N-1)]$, and thus for $h \neq 0$ and large $N$, or more generally for $hN \gg 1$, the perturbation matrix element $P^{\prime}_{N,1}$ takes extremely large values, and Eq.(13) becomes invalid even for extremely low values of $\epsilon$, of the order $\sim \exp(-hN)$. This result agrees with previous studies showing the strong dependence of the spectrum of the Hatano-Nelson Hamiltonian on boundary conditions \cite{R57,R63,R64} and  is a clear signature that in a non-Hermitian Hamiltonian near an EP a small change of a control parameter can induce a comparatively much larger change in its energy spectrum. For example, let us consider a uniform tight-binding chain with nearest-neighbor hopping rate $\kappa$ and open boundary conditions, defined by the Hamiltonian $H_0$ given in Eq.(5), and let assume that the perturbation $\epsilon P$ describes a small (Hermitian) coupling between the edge sites $n=1$ and $n=N$ of the chain, i.e. let us assume 
\begin{equation}
P_{n,m}= \delta_{n,1}\delta_{m,N}+\delta_{n,N}\delta_{m,1}.
\end{equation}
 This can be readily obtained by deforming a linear chain so that to weakly couple the edge sites, as shown in Fig.1. The perturbed Hamiltonian reads explicitly
\begin{equation}
 {H}_2=\left(
 \begin{array}{ccccccc}
 0 & \kappa \exp(h) & 0 &  ... & 0 & \epsilon \\
 \kappa \exp(-h) & 0 & \kappa \exp(h) &  ... & 0 & 0 \\
 0 & \kappa \exp(-h) & 0 &  ... & 0 & 0 \\
 ... & ... & ... & ...& ... & ... \\
 \epsilon & 0 & 0 & ... & \kappa \exp(-h)& 0 
 \end{array}
 \right).
 \end{equation}
For $\epsilon=0$, i.e. in the absence of the perturbation, the energy spectrum of $H_2$ is real, independent of $h$ and given by
\begin{equation}
E_n=2 \kappa \cos[2 \pi n/(N+1)]
\end{equation}
($n=1,2,...,N$), with corresponding eigenfunctions 
\begin{equation}
({\bf v}_n)_l=\sqrt{\frac{2}{N+1}} \sin \left( \frac{nl \pi}{N+1}\right) \exp(-lh).
\end{equation}
In the presence of the long-range perturbation (14), the energies are modified according to Eq.(13) as follows
\begin{eqnarray}
E^{\prime}_n &  \simeq  & E_n+{\frac{2 \epsilon}{N+1}} \sin^2 \left( \frac{n \pi }{N+1} \right)   \nonumber \\
 &  \times & \left\{ \exp[(N-1)h]+\exp[-(N-1) h] \right\} .
\end{eqnarray}
The perturbative analysis would predict the energy spectrum to remain real, however it is clear that  in long chains even for small $\epsilon$ [of the order of $\sim \kappa \exp(-hN)$] the correction of the energy ceases to be small and the perturbative analysis is expected to fail even for $\epsilon$ much smaller than the smallest hopping rate $\kappa_2=\kappa \exp(-h)$. {The exact eigenvalues $E^{\prime}_n$ (energy spectrum) of the matrix $H_2$ can be computed from the roots of a self-inversive polynomial, as shown in the Appendix A. In particular, as $\epsilon$ is increased above a critical value $\epsilon_c$, the energy spectrum ceases to be real and pairs of real energies coalesce and bifurcate into complex conjugate energies via an EP. For $\cosh[(N-1)h] \gg 1$, the critical value $\epsilon_c$ of perturbation strength takes the simple form $\epsilon_{c} \simeq \kappa / \{ 2 \cosh [(N-1) h] \}$.}  As an example, Fig.2 shows the numerically-computed exact energy spectrum of the perturbed Hamiltonian $H_2$ in a lattice comprising $N=50$ sites for a few increasing values of $\epsilon$ and for   $h=0$ [Hermitian limit, Fig.2(a)], $h=0.1$ [Figs.2(b)] and $h=0.2$ [Fig.2(c)]. As one can see, for a non-vanishing imaginary gauge field even for small values of $\epsilon$ the energy spectrum of the perturbed Hamiltonian $H_2$ is strongly modified as compared to  the spectrum of the unperturbed one $H_1$, and rapidly bifurcates into complex energies as $\epsilon$ is increased {above $\epsilon_c$. A typical bifurcation scenario is shown in Fig.2(d), where the loci of eigenvalues $E^{\prime}$ in complex plane are depicted for increasing the perturbation strength, from below to above the critical value $\epsilon_c$. As one can see, as $\epsilon$ in increased couples of real energies coalesce via an EP and bifurcate into complex conjugate ones.}

\begin{figure}
  \includegraphics[width=\columnwidth]{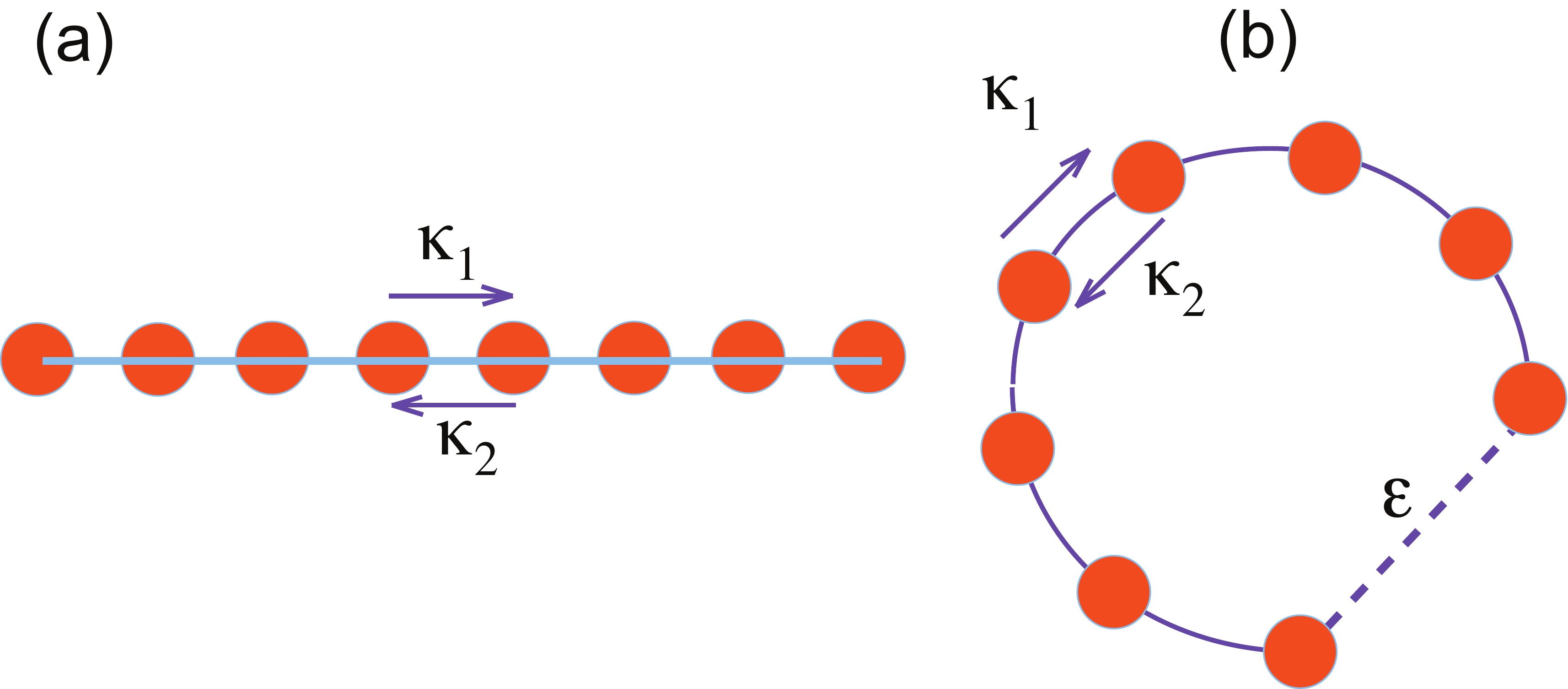}%
  \caption{\label{<label name>}\col 
   (a) Schematic of a linear chain with open boundary conditions and asymmetric hopping rates $\kappa_1=\kappa \exp(h)$ and $\kappa_2=\kappa \exp(-h)$. (b) A long-range perturbation can be obtained by deforming the linear chain into an arc so as the edge sites are weakly coupled by a strength $\epsilon \ll \kappa$.}
\end{figure}

\begin{figure}
  \includegraphics[width=\columnwidth]{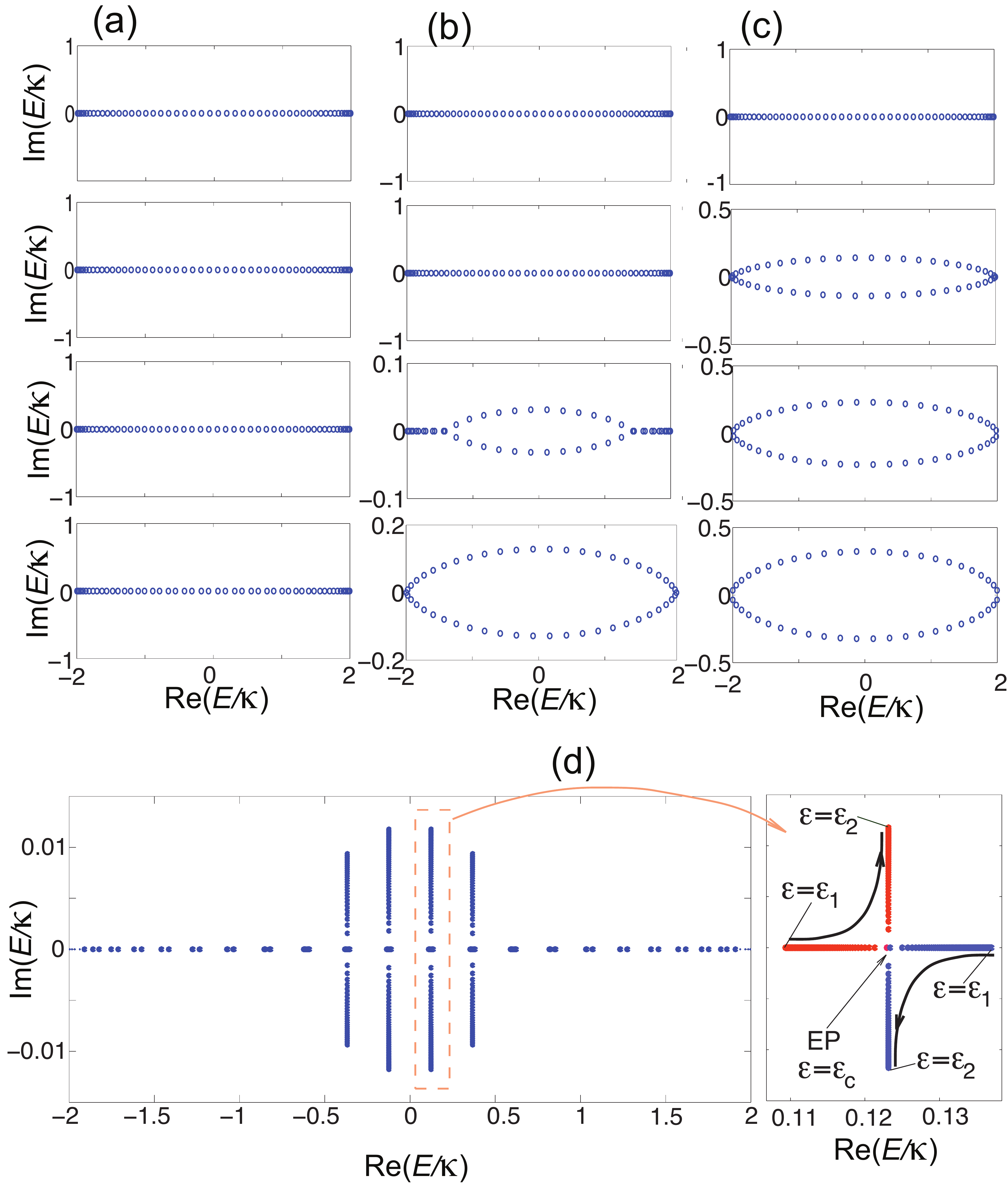}%
  \caption{\label{<label name>}\col 
   Energy spectrum of the matrix $H_2$, defined by Eq.(15), for a few increasing values of the imaginary gauge phase $h$: (a) $h=0$ (Hermitian limit), (b) $h=0.1$, and (c) $h=0.2$. The perturbation strength $\epsilon / \kappa $ is (from top to bottom) 0, 0.001, 0.01 and 0.1. {(d) Loci of the energies $E$ of $H_2$ in complex plane for $h=0.1$, $N=50$ and for increasing values of the perturbation strength $\epsilon$, from $\epsilon=\epsilon_1=0.007 \kappa$ to $\epsilon=\epsilon_2=0.0078 \kappa$. The critical perturbation strength is $\epsilon_c= \kappa/ \{ 2 \cosh[(N-1)h]\} \simeq 0.0074 \kappa$. The right panel in (d) shows the detailed behavior of a pair of energies that coalesce via an EP and bifurcate from real to complex conjugate values as $\epsilon$ is increased (arrows indicate the path of eigenvalues as $\epsilon$ is increased).}}
\end{figure}

\section{Dynamical stability under perturbations: fidelity decay}
The enhanced sensitivity of energy spectrum of the pseudo-Hermitian Hamiltonian $H_1$ to a perturbation, as compared to the Hermitian case, corresponds rather generally to a faster deviation of the system temporal dynamics as indicated by a faster decay of fidelity. Let us consider an initial state ${\boldsymbol \psi}_0$ at time $t=0$, and let us consider its temporal evolution under the unperturbed Hamiltonian $H_1$ and the perturbed one $H_2=H_1+\epsilon P$. After setting ${\boldsymbol \psi}_1(t)= \exp(-i H_1 t) {\boldsymbol \psi}_0$ and ${\boldsymbol \psi}_2(t)= \exp(-i H_2 t) {\boldsymbol \psi}_0$, the deviation of the dynamics induced by the perturbation is measured by the fidelity \cite{R46}
\begin{equation}
\mathcal{F}(t)=\frac{| \langle {\boldsymbol \psi}_2(t) | {\boldsymbol \psi}_1(t) \rangle |^2}{\langle {\boldsymbol \psi}_1(t) | {\boldsymbol \psi}_1(t) \rangle \langle {\boldsymbol \psi}_2(t) | {\boldsymbol \psi}_2(t) \rangle}
\end{equation}
where the denominator is introduced because of the non-unitary dynamics. Note that $\mathcal{F}(t) \leq 1$ and $\mathcal{F}(t)=1$ if and only if the state vectors ${\boldsymbol \psi}_1(t)$ and ${\boldsymbol \psi}_2(t)$ are parallel. A special case corresponds to the initial state ${\boldsymbol \psi}_0$ being prepared in an eigenstate of the unperturbed Hamiltonian, such as the ground (equilibrium)  state: in this case the fidelity measures the deviations of the dynamics after a sudden quench of the Hamiltonian from $H_1$ to $H_2$.
Intuitively, one would expect that the fidelity decay should be faster as the non-Hermitian gauge phase $h$ is increased because of the stronger deformation of the energy spectrum for increasing values of $h$ (see Fig.2). Such a behavior is observed, for instance, in quench dynamics, where the initial state ${\boldsymbol \psi}_0$ is an eigenstate of the unperturbed Hamiltonian [see Fig.4(a) to be commented below]. However, what happens when system is initially prepared in a state which is not an equilibrium state, i.e. in a superposition of eigenstates of $H_1$? We show here that, for long-range perturbations and for initial excitations that are confined very far from the region of skin edge eigenstates, a very counterintuitive behavior can be observed as a result of asymmetric hopping and non-Hermitian skin effect: the fidelity decay can be {\em decelerated} (rather than accelerated), at least transiently, by increasing the imaginary gauge field $h$, i.e. by bringing the system closer to an EP. A simple physical picture of such a counterintuitive behavior can be gained as follows. Let us consider the time-dependent Schr\"odinger equation for the unperturbed and perturbed Hamiltonians
\begin{eqnarray}
i \partial_t {\boldsymbol \psi}_1 & = & H_1 {\boldsymbol \psi}_1  \\
 i \partial_t {\boldsymbol \psi}_2 & = & H_2 {\boldsymbol \psi}_2
\end{eqnarray}
with ${\boldsymbol \psi}_1(0)={\boldsymbol \psi}_2(0)={\boldsymbol \psi}_0$. After setting
\begin{equation}
{\boldsymbol \psi}_2(t)={\boldsymbol \psi}_1(t)+ \delta {\boldsymbol \psi}(t)
\end{equation}
the evolution equation for the variation $\delta {\boldsymbol \psi}$ reads
\begin{equation}
i \partial_t \delta {\boldsymbol \psi}=H_2 \delta {\boldsymbol \psi} +\epsilon P {\boldsymbol \psi}_1
\end{equation}
with $\delta {\boldsymbol \psi}(0)=0$. If we expand the fidelity in power series of the variation $\delta {\boldsymbol \psi}(t)$, up to second-order expansion one obtains
\begin{equation}
\mathcal{F}(t) \simeq 1-\frac{\langle \delta {\boldsymbol \psi }| \delta {\boldsymbol \psi } \rangle}{\langle {\boldsymbol \psi }_1 |  {\boldsymbol \psi } _1 \rangle}+\frac{| \langle \delta {\boldsymbol \psi }|  {\boldsymbol \psi }_1 \rangle |^2}{\langle {\boldsymbol \psi }_1 |  {\boldsymbol \psi } _1 \rangle ^2 }.
\end{equation}
\begin{figure}
  \includegraphics[width=\columnwidth]{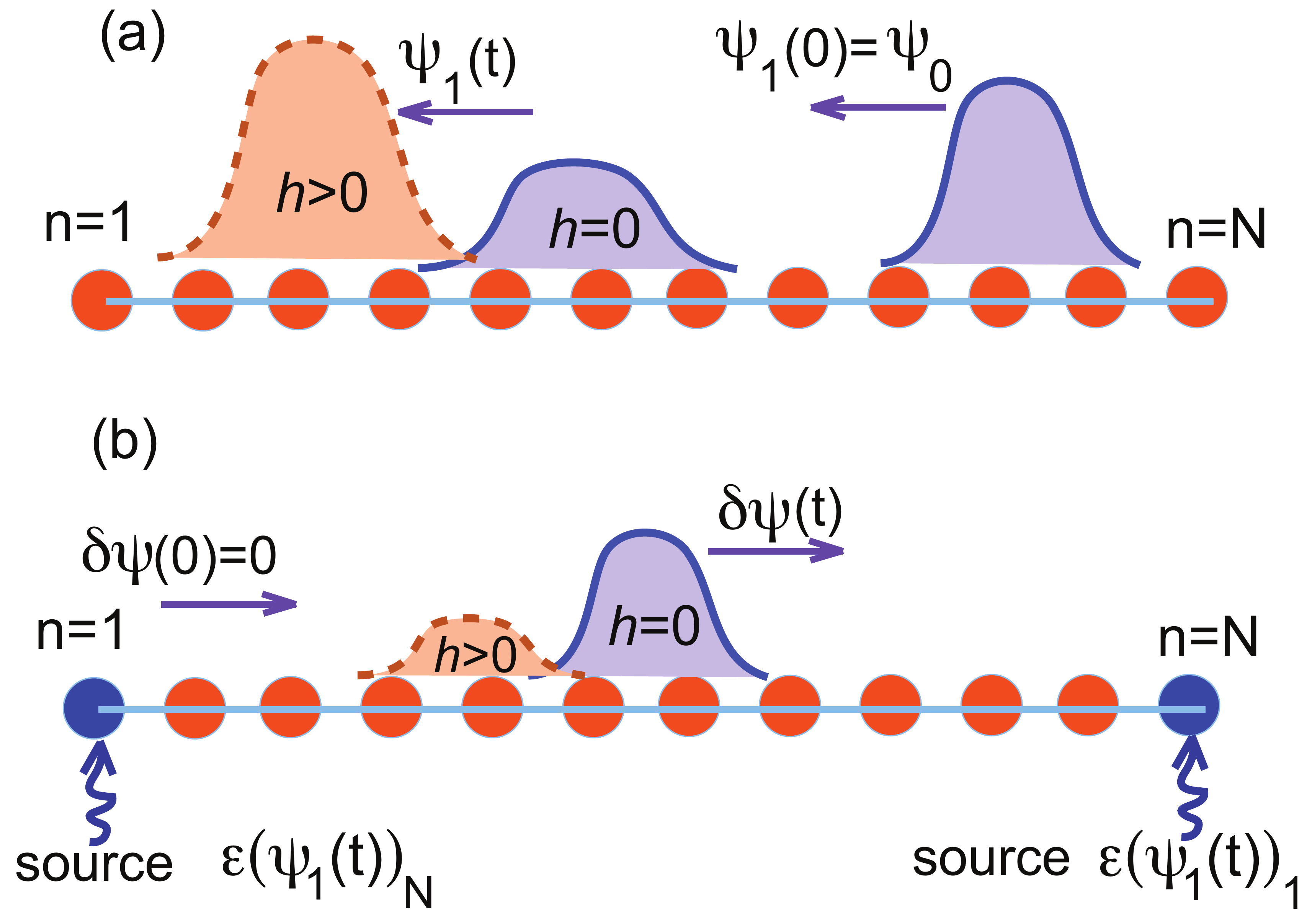}%
  \caption{\label{<label name>}\col 
   (a) Schematic of wave packet propagation in the unperturbed chain, described by the Hamiltonian $H_1$, for initial excitation ${\boldsymbol \psi}_1(0)={\boldsymbol \psi}_0$ localized at the right edge of the chain. The propagated wave packet ${\boldsymbol \psi}_1(t)$  at successive time $t$ is shown for a vanishing gauge field $h=0$ (Hermitian limit) and for $h>0$ by solid and dashed curves, respectively. Arrows indicate the propagation direction of the wave packets. In the non-Hermitian case the wave packet is amplified and propagates faster. (b) The long-range perturbation $P$ [Eq.(14)] changes the evolution of the wave packet according to ${\boldsymbol \psi}_2(t)={\boldsymbol \psi}_1(t)+\delta {\boldsymbol \psi}(t)$. The  evolution of the wave packet correction $\delta {\boldsymbol \psi}(t)$ is governed by the same chain in (a) but with  two sources at the two edge sites. The source on the right edge side is negligible, so that  $\delta {\boldsymbol \psi}$ is created on the left edge and propagates forward in the chain. A schematic behavior of $\delta {\boldsymbol \psi}(t)$ for a vanishing ($h=0$) and for a non-vanishing ($h>0$) gauge field is shown by solid and dashed curves, respectively. In the non-Hermitian case the wave packet correction $\delta {\boldsymbol \psi}(t)$ propagates slower and is attenuated.}
\end{figure}
For the sake of definiteness, let us assume the long-range perturbation defined by Eq.(14), however a similar analysis holds for any perturbation with matrix elements $P_{n,m}$ non vanishing only for $|n-m|$ large enough. Let us also assume that the initial state ${\boldsymbol \psi }_0$ is localized on the right side of the chain, so that $({\boldsymbol \psi }_0)_n=0$ for small values of index $n$; see Fig.3(a) for a schematic. To provide an estimation of the fidelity $\mathcal{F}$, it is sufficient to provide a qualitative behavior of the solutions ${\boldsymbol \psi}_1$ and $ \delta {\boldsymbol \psi}$ of Eqs.(20) and (23). 
Clearly, in the unperturbed system with Hamiltonian $H_1$ the initial excitation ${\boldsymbol \psi}_0$, localized on the few right edge sites of the chain, spreads and propagates backward with a speed which is ultimately limited by the largest hopping rate; see Fig.3(a) for a schematic. To provide a qualitative behavior of the solution $\delta {\boldsymbol \psi}(t)$, note that, since $\delta {\boldsymbol \psi}(0)=0$, in the early stage of the dynamics one can take  $H_{2}=H_1+ \epsilon P \simeq H_1$ on the right-hand-side of Eq.(23), i.e one can assume
\begin{equation}
i \partial_t \delta {\boldsymbol \psi} \simeq H_1 \delta {\boldsymbol \psi} +\epsilon P {\boldsymbol \psi}_1
\end{equation}
Since $(P {\boldsymbol \psi}_1)_n=({\boldsymbol \psi}_1)_N \delta_{n,1}+({\boldsymbol \psi}_1)_1 \delta_{n,N}$ , the solution $\delta {\boldsymbol \psi}(t)$ is basically determined by the propagation in the same chain described by $H_1$ but with two sources $\epsilon ({\boldsymbol \psi}_1)_N$ and $\epsilon ( {\boldsymbol \psi}_1)_1 $ that create the excitation. The two sources are placed at the left and right edges sites $n=1$ and $n=N$, respectively; see Fig.3(b) for a schematic. Since the wave packet ${\boldsymbol \psi}_1(t)$ is initially localized near the $n=N$ edge of the chain and propagates at a finite speed along the chain, in the early stage of the dynamics one can assume $({\boldsymbol \psi}_1)_1 \simeq 0$, so that the only source for $\delta {\boldsymbol \psi}(t)$ is located at the left edge. This means that in the early stage of the dynamics, i.e. until the initial excitation ${\boldsymbol \psi}_1(t)$ has not spread from the right  to the left boundaries of the chain and the correction $ \delta {\boldsymbol \psi}(t)$ has not spread from the left to the right extremes of the chain, one can assume $\langle {\boldsymbol \psi}_1(t) | \delta {\boldsymbol \psi}(t) \rangle \simeq 0$ in Eq.(24), so that
\begin{equation}
\mathcal{F}(t) \simeq 1-\frac{\langle \delta {\boldsymbol \psi }| \delta {\boldsymbol \psi } \rangle}{\langle {\boldsymbol \psi }_1 |  {\boldsymbol \psi } _1 \rangle}.
\end{equation}
In the Hermitian limit $h=0$, the norm $\langle {\boldsymbol \psi }_1 |  {\boldsymbol \psi } _1 \rangle=\langle {\boldsymbol \psi }_0 |  {\boldsymbol \psi } _0 \rangle=1$ is conserved, whereas $\langle \delta {\boldsymbol \psi }| \delta {\boldsymbol \psi } \rangle$ increases from zero because of the source term in Eq.(25): hence the fidelity decays like $\mathcal{F}(t)=1-{\langle \delta {\boldsymbol \psi }| \delta {\boldsymbol \psi } \rangle}$. For a non-vanishing imaginary gauge field $h$, the wave packet ${\boldsymbol \psi}_1(t)$ propagates faster and it is exponentially {\em amplified} while propagating backward in the chain \cite{R53,R54}. Hence the norm $\langle {\boldsymbol \psi }_1 |  {\boldsymbol \psi } _1 \rangle$ is not conserved and turns out to be an almost exponentially-increasing function of time. Likewise, the wave packet $\delta {\boldsymbol \psi}(t)$ created by the source on the left edge is exponentially {\em attenuated} while propagating forward along the chain \cite{R53,R54}, so that the norm  $\langle \delta {\boldsymbol \psi }| \delta {\boldsymbol \psi } \rangle$ for $h>0$ takes smaller values as compared to the ones for $h=0$. This means that the fidelity $\mathcal{F}(t)$ is expected to {\em increase} when $h$ is increased from zero, i.e. an increase of the imaginary gauge field leads to a {\em deceleration} (rather than an acceleration) of the fidelity decay, despite the energy spectrum of the Hamiltonian $H_2$ undergoes a stronger deformation as $h$ is increased.\\ We checked the predictions of the theoretical analysis by numerically computing the fidelity decay for the nearest-neighbor tight-binding Hamiltonians $H_1$ and $H_2$, defined by Eqs.(6) and (15) with $N=50$ sites,  for two different initial conditions ${\boldsymbol \psi}_0$. The numerical results are shown in Fig.4 for a vanishing ($h=0$) and non-vanishing ($h=0.3$) gauge field. In Fig.4(a) the initial state ${\boldsymbol \psi}_0$ is the eigenstate ${\bf v}_1$ of $H_1$, given by Eq.(17) with $n=1$. Clearly, a non-vanishing imaginary gauge field accelerates the decay of fidelity [compare left and right panels in Fig.4(a)]. This is an expected result because the gauge field effectively enhances the strength of perturbation and greatly modify the eigen-energies of the perturbed Hamiltonian $H_2$ as compared to the spectrum of $H_1$, as discussed in the previous section. Figure 4(b) shows the typical decay behavior of the fidelity for the initial state $({\boldsymbol \psi}_0)_n=\delta_{n,N-1}$, corresponding to initial excitation of a site close to the edge right site of the chain. In this case, one can see that in the early stage the fidelity decay is {\em decelerated} (rather than accelerated) by a non-vanishing imaginary gauge field $h$, until an abrupt drop of the fidelity is observed at the time $ t_1 \simeq 24 / \kappa \simeq N/v_g$, corresponding  to the transit time of the excitation along the chain (the group velocity being given by $v_g= 2 \kappa \cosh (h)$ \cite{R53}). The time $t_1$ of the fidelity drop can be increased by increasing the length $N$ of the chain.
\begin{figure}
  \includegraphics[width=\columnwidth]{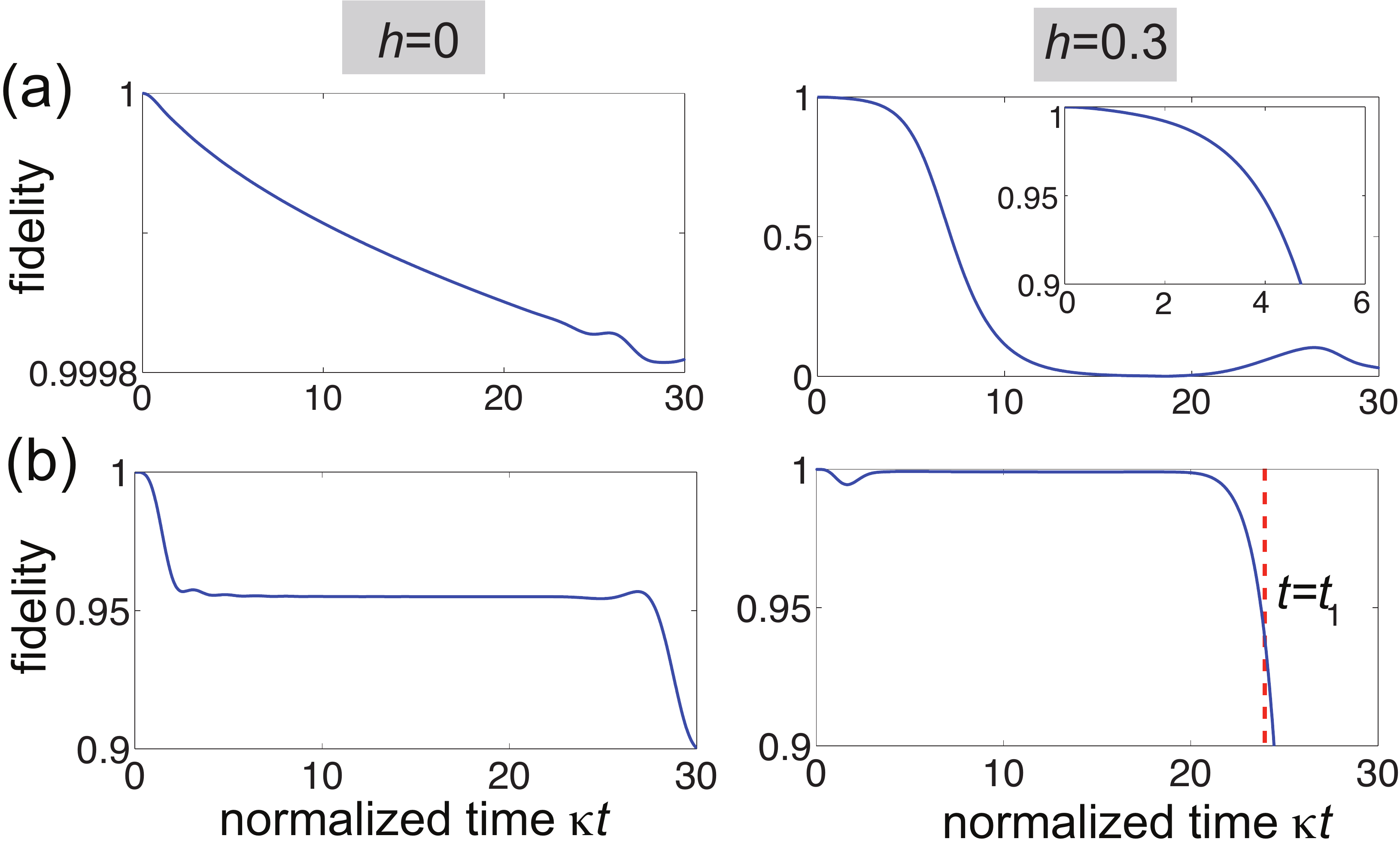}%
  \caption{\label{<label name>}\col 
   Numerically-computed behavior of fidelity decay for the tight-binding Hamiltonians $H_1$ and $H_2$ defined by Eqs.(6) and (15) with $N=50$, $\epsilon=0.3 \kappa$ and $h=0$ (Hermitian limit, left panels) and $h=0.3$ (right panels). (a) and (b) correspond to the two different initial states ${\boldsymbol \psi}_0$ discussed in the main text. The inset in the right panel of (a) is an enlargement of the fidelity decay in the early stage of the dynamics. In (b) a slowing down of fidelity decay, up to the time $t_1 \simeq N/v_g$ (the transit time of excitation all along the chain), is clearly observed in the non-Hermitian case.}
\end{figure}
\begin{figure}
  \includegraphics[width=\columnwidth]{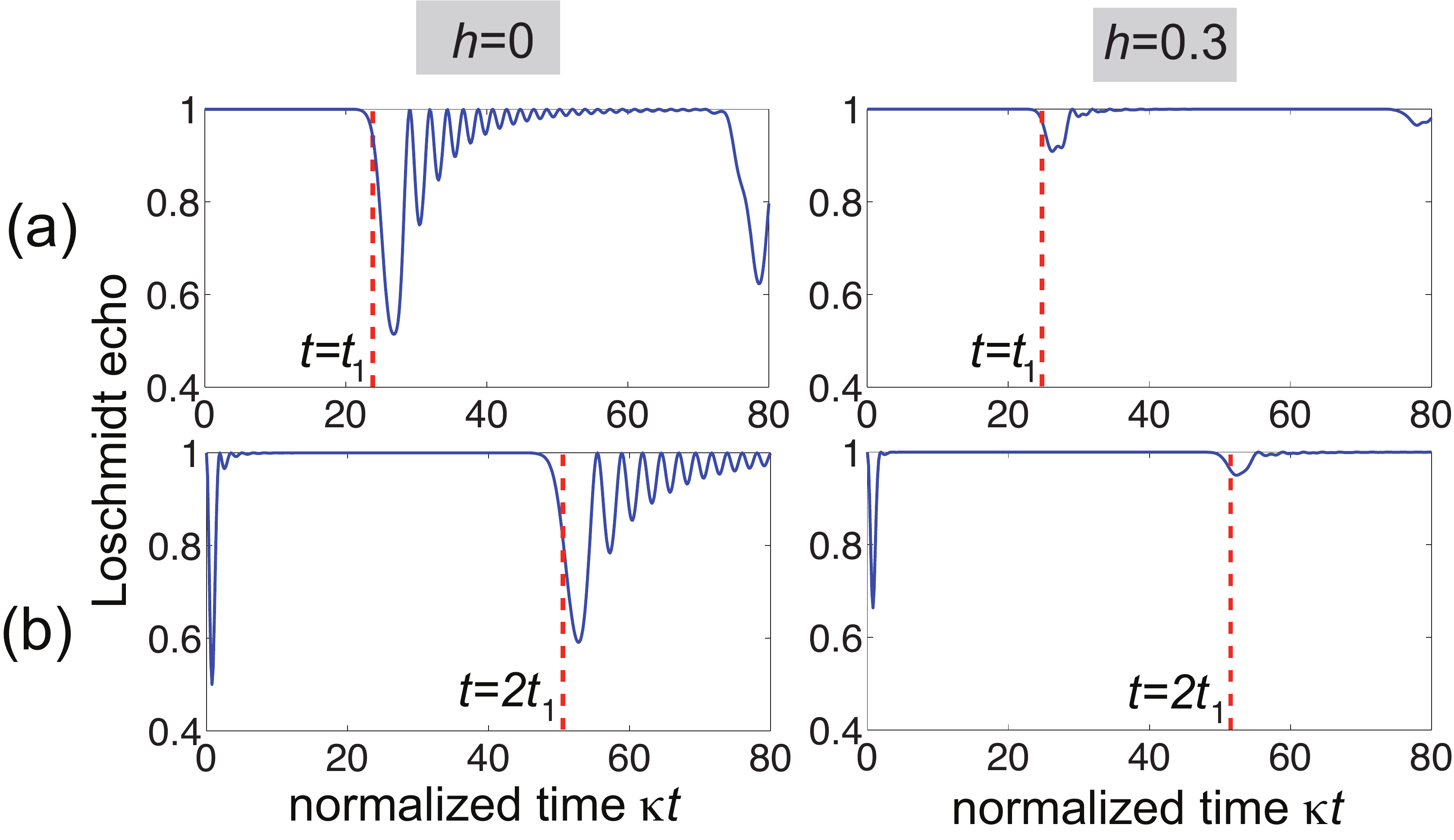}%
  \caption{\label{<label name>}\col 
   Numerically-computed behavior of Loschmidt echo $M(t)$ for the pseudo Hermitian  Hamiltonian (6) with open boundary conditions, as obtained by sudden phase slip of the wave function amplitudes $({\boldsymbol \psi})_n$ at time $t$ according to Eq.(31). The number of lattice sites is $N=50$. The system is initially prepared in the state $({\boldsymbol \psi}_0)_n=\delta_{n,1}$, corresponding to the single-site excitation of the left edge site of the lattice. (a) and (b) correspond to different position $n_0=N$ and $n_0=1$ of the phase defect, respectively. Left and right panels show the Loschmidt echo for the two different values $h=0$ (Hermitian limit) and $h=0.3$ of the imaginary gauge field. The vertical dashed lines in (a) depict the time interval $t_1$ required for the initial excitation to traverse the entire chain, whereas in (b) the round-trip traversal time $2t_1$ is highlighted.}
\end{figure}
\begin{figure}
  \includegraphics[width=\columnwidth]{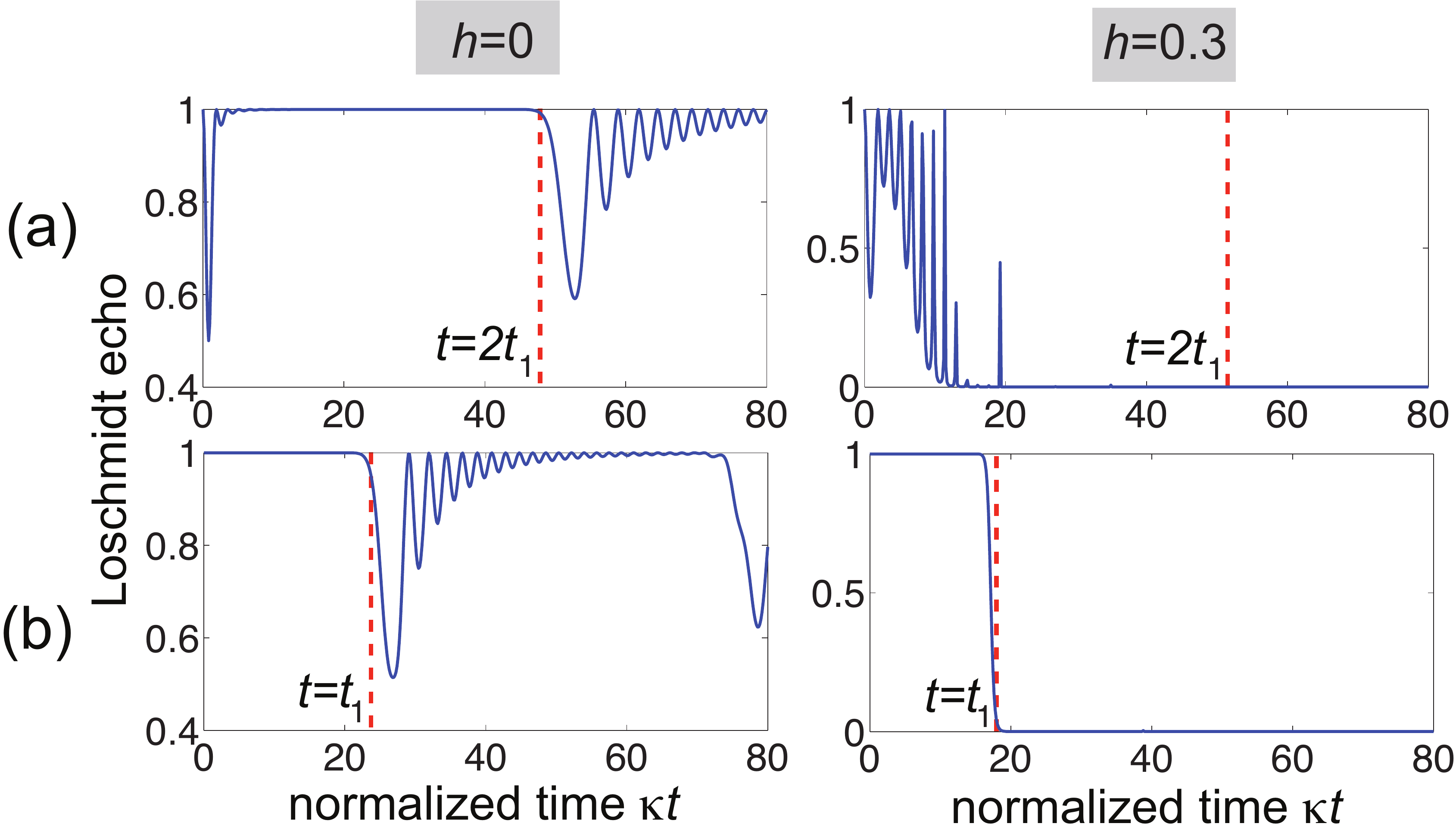}%
  \caption{\label{<label name>}\col 
   Same as Fig.5, but for the initial state $({\boldsymbol \psi}_0)_n=\delta_{n,N}$, corresponding to the single-site excitation of the right edge site of the lattice.}
\end{figure}

\section{Imperfect time reversal and Loschmidt echo}
The reversibility of dynamical classical and quantum systems is generally  captured by Loschmidt echo, which  provides a quantitative measure of the sensitivity of perturbations of the backward temporal evolution
of the system \cite{R46,R48}. In typical Loschmidt echo setups, an initial state ${\boldsymbol \psi}_0$ is forward propagated for a given time $t$ with the Hamiltonian $H_1$ and then imperfectly time reversed and propagated backward by the Hamiltonain $H_2 \simeq H_1$, resulting in a final state 
\begin{equation}
{\boldsymbol \psi}_f=\exp(iH_2 t) \exp(-i H_1 t) {\boldsymbol \psi}_0.
\end{equation} 
The Loschmidt echo $M(t)$ is defined as the overlap between the original state ${\boldsymbol \psi}_0$ and the final state ${\boldsymbol \psi}_f$ after imperfect time reversal, i.e. 
\begin{equation}
M(t)=\frac{\left| \langle {\boldsymbol \psi}_0 | {\boldsymbol \psi}_f \rangle \right|^2}{ \langle {\boldsymbol \psi}_f | {\boldsymbol \psi}_f \rangle \langle {\boldsymbol \psi}_0 | {\boldsymbol \psi}_0 \rangle}
\end{equation}
with $M(t) \leq 1$ and $M(t)=1$ only for perfect time reversal. { Clearly, in the Hermitian case the norm is conserved and $\hat{H_2}^{\dag}=\hat{H_2}$, so that from Eqs.(19), (27) and (28) it follows that $\mathcal{F}(t)=M(t)$, i.e. the Loschmidt echo and fidelity do coincide. This means that the sensitivity of the dynamical evolution to perturbations of the Hamiltonian can be obtained either from the imperfect time reversal of the dynamics of a single system or from a comparison of the different dynamical evolutions of two  systems prepared in the same initial state but evolving under different (unperturbed and perturbed) Hamiltonians. When considering the non-Hermitian case, the two quantities $\mathcal{F}(t)$ and $M(t)$ do not coincide anymore, since time reversal requires to change $^{\prime}$gain$^{\prime}$ with $^{\prime}$loss$^{\prime}$ terms in the Hamiltonian. Nonetheless, both quantities provide a measure of sensitivity of perturbations in the dynamical evolution of a non-Hermitian Hamiltonian in the two different physical settings.}  A main result that we wish to highlight in this section is that, despite the major sensitivity to perturbations, for certain initial states an imaginary gauge field $h$ can result in an {\em increase} (rather 
than a decrease) of the Loschmidt echo after imperfect time reversal, as compared to the Hermitian limit $h=0$. To illustrate such a behavior, let us focus our attention to a tight-binding chain with nearest-neighbor hopping rate $\kappa$, defined by the Hatano-Nelson Hamiltonian (6). In this case time reversal, after forward propagation with the Hamiltonian $H_1$ for a time interval $t$, is simply obtained by flipping the sign of the hopping rate $\kappa$ at time $t$ . Note that, since the gauge field is imaginary, time reversal does not require to flip the sign of $h$ in the Hamiltonian. Time reversal for this kind of Hamiltonian in the Hermitian limit $h=0$ has been suggested and experimentally demonstrated in a series of recent works for cold atoms in optical lattices and  for photons in evanescently coupled optical waveguide arrays \cite{R71,R72,R73,R74,R75,R76,R77}. As discussed in such previous works, the sign flip of the hopping amplitude $\kappa$ is obtained by fast half-cycle Bloch oscillations or by imprinting a sudden phase shift to adjacent amplitudes. In the presence of an imaginary gauge field, time reversal can be obtained using the same method. Let us assume, for example, that imperfect time reversal is obtained by imprinting a sudden site-dependent phase shift $\phi_n$ to the amplitudes $({\boldsymbol \psi})_n$, described by the operator $\mathcal{P}$ as follows
\begin{equation}
{\mathcal P} ({\boldsymbol \psi})_n=({\boldsymbol \psi})_n \exp(i \phi_n).
\end{equation}
The final state of the system is then given by
\begin{equation}
{\boldsymbol \psi}_f= \exp(-i H_1 t) {\mathcal{P}} \exp(-i H_1 t) {\boldsymbol \psi}_0
\end{equation}
For $\phi_n= \pi$, one has perfect time reversal since in this case $ \exp(-i H_1 t) {\mathcal{P}}= \exp(i H_1 t)$. In practice, especially for a large number of sites $N$, the phase shifts imprinted to the amplitudes $({\boldsymbol \psi})_n$ can deviate from the target $\pi$ value, resulting in an effective imperfect time reversal and lowing of the Loschmidt echo. For example, let us assume that a wrong phase, equal to $\pi/2$ (rather than $\pi$), is impressed at the lattice site $n=n_0$, i.e. let us assume
\begin{equation}
\phi_n= \pi -(\pi/2) \delta_{n,n_0}.
\end{equation}
Let us consider initial state excitation ${\boldsymbol \psi}_0$ in a single site of the chain located at the left edge, i.e. $({\boldsymbol \psi}_0)_n=\delta_{n,1}$. Figure 5 shows a typical behavior of the Loschmidt echo $M(t)$ in the Hermitian ($h=0$) and pseudo-Hermitian ($h>0$) cases corresponding to two different values of $n_0$, i.e. of position of the phase shift defect in the chain. Clearly, in both cases the Loschmidt echo turns out to be larger for a non-vanishing imaginary gauge field. Such a result can be physically explained on the basis of the non-Hermitian skin effect in the Hatano-Nelson model with open boundary conditions. Let us consider, for instance, the case of Fig.5(a), corresponding to the phase defect at the right site $n_0=N$ of the chain. In the Hermitian limit $h=0$ [left panel of Fig.5(a)] the Loschmidt echo remains very close to one for times $t$ smaller than the characteristic time $t \simeq t_1$ required for the initial excitation to reach the right boundary of the chain: in fact, for $t<t_1$ the initial excitation spreads along the chain but it is refocused before it can reach the right edge site, and thus the presence of the phase defect at the right edge site is not probed. On the other hand, for times $t$ larger than $t_1$ the excitation reaches the right edge site and the time reversal is thus imperfect, as one can see from the rather abrupt drop of the Loschmidt echo. Let us consider now the non-Hermitian case [right panel in Fig.5(a)].
Owing to the non-Hermitian skin effect, a positive imaginary gauge field $h$ pushes the excitation toward the left edge site, so that  even for times larger than $t_1$ the right edge state with the phase defect is only weakly probed by the excitation, and thus the time reversal process is less sensitive to the phase defect. This is clearly shown in the right panel of Fig.5(a), where the drop of the Loschmidt echo near the time $t \sim t_1$ is much smaller than in the Hermitian case and the interference (oscillatory) fringes observed in the Hermitian case are washed out. A similar simple physical explanation can be given for the increase of Loschmidt echo observed by a non-vanishing imaginary gauge field for the case of Fig.5(b), where the phase defect is placed at the left edge site of the chain. It should be noted that for a different initial condition the Loschmidt echo can be degraded by the imaginary gauge phase. For example, if the system is initially excited in a single site (as in Fig.5) but {\em on the right edge}, such as $({\boldsymbol \psi})_n=\delta_{n,N}$, the imagery gauge field degrades the Loschmidt echo (see Fig.6). The reason thereof is that in this case the non-Hermitian skin effect amplifies the excitation that probes the phase defect site, so as the imperfection of the time reversal process is more effective when $h>0$. In some sense, one can say that the imaginary gauge field introduces a kind of {\em squeezing} in the dynamical stability of the system: the imaginary gauge field enhances time reversibility for some initial conditions (those with stronger localization at the left edge of the lattice, where the gauge field pushes the excitation) but degrades time reversibility for other initial conditions (those with stronger localization at the right edge).

 \section{Conclusion.} The dynamical behaviors of non-Hermitian classical and open quantum systems near exceptional points are attracting a great interest in recent years, featuring some rather unique effects with no counterpart in Hermitian systems. A rather general result is that, for a non-Hermitian Hamiltonian that depends on a control parameter $h$ and shows an EP, the dynamical behavior of the system becomes much more sensitive to perturbations and/or initial conditions as the system is operated closer to the EP. The stability of the dynamical behavior of Hermitian systems is usually described by Loschmidt echo and fidelity decay. Such quantities are commonly introduced in problems related to quantum chaos, ergodicity, decoherence, non-equilibrium dynamics of many body systems, etc.  However, so far they 
 have been rarely considered as a measure of dynamical stability in non-Hermitian systems \cite{R38uff,R38tris}. Here we have used fidelity and Loschmidt echo to investigate the dynamical stability of certain non-Hermitian models and 
 disclosed a rather surprising result. Owing to the strong sensitivity of non-Hermitian Hamiltonians near an EP, one would naively expect a less stable dynamical behavior signaled by degradation of Loschmidt echo and faster fidelity decay dynamics. Contrary to such an expectation, in this work we have shown that, for a class of pseudo-Hermitian Hamiltonians, a system operated closer to an EP can show a {\em deceleration} (rather than an acceleration) of the fidelity decay and an enhanced Loschmidt echo under a broad class of perturbations and initial excitations. We have illustrated such a counterintuitive effect by considering non-Hermitian transport in the Hatano-Nelson model, where an imaginary gauge field introduces asymmetric transport in the lattice, and provided a simple physical explanation of the main results on the basis of the so-called non-Hermitian skin effect. The present study discloses unexpected features in the dynamical behavior of certain non-Hermitian systems, and  should motivate further investigations on dynamical stability of non-Hermitian classical and quantum systems. For example, it would be interesting to investigate dynamical stability in other non-Hermitian models, such as $\mathcal{PT}$-symmetric models, non-Hermitian many-particle systems, where effects such as particle statistics and correlations should play a major role \cite{fine}, dynamical stability in presence of two or more EPs in parameter space, and the interplay between topology and dynamical stability \cite{R38uff}.

  \appendix
  \renewcommand{\theequation}{A.\arabic{equation}}
 \setcounter{equation}{0}

 \section{Appendix}
{ In this Appendix we calculate the exact form of the eigenvalues and corresponding eigenvectors of the matrix $H_2$ given by Eq.(15) in the main text. Let $\mathbf{v}^{\prime}=(c_1,c_2,...,c_N)^T$ be an eigenvector of $H_2$ with eigenvalue $E^{\prime}$, i.e.
\begin{equation}
H_2 \mathbf{v}^{\prime}=E^{\prime} \mathbf{v}^{\prime}.
\end{equation}
 We look for a solution of the eigenvector elements $c_n$ in the form of counter-propagating waves, i.e. we make the Ansatz
\begin{equation}
c_n=\{ A \exp [iq(n-1)]+B \exp[-iq(n-1)] \} \exp[-h(n-1)]
\end{equation}
($n=1,2,3,...,N$). The corresponding eigenvalue is given by
\begin{equation}
E^{\prime}=2 \kappa \cos (q)
\end{equation}
as one can readily prove after substitution of Eq.(A.2) into Eq.(A.1) and taking $n=2,3,...,N-1$. In the above equations, $q$ is a complex parameter that should be determined from a solvability condition, and $A$,$B$ are the amplitudes of the two counter-propagating waves in the lattice. By writing down Eq.(A.1) for $n=1$ and $n=N$ and using Eqs.(A.2) and (A.3), the following homogeneous equations for the amplitudes $A$ and $B$ are obtained
\begin{eqnarray}
 \left( \frac{\kappa}{y} -  {\epsilon \rho}{ y^{N-1}} \right) A +\left( \kappa y- \frac{\epsilon \rho}{y^{N-1}} \right) B=0\\
  \left( \kappa y^N -  \frac{\epsilon}{\rho} \right) A +\left( \frac{\kappa}{y^{N}} -\frac{\epsilon}{\rho} \right) B=0
\end{eqnarray}
where we have set
\begin{eqnarray}
y & \equiv & \exp(iq) \\
 \rho & \equiv & \exp[-h(N-1)].
\end{eqnarray}
}
A non-vanishing solution for $A$ and $B$ requires the vanishing of the determinant in Eqs.(A.4) and (A.5). This solvability condition implies that $y$ is a root of the following polynomial $P(y)$ of order $2(N+1)$:
\begin{eqnarray}
P(y) & = & y^{(2N+2)}-\frac{\epsilon^2}{\kappa^2}y^{2N}-\frac{2 \epsilon}{\kappa}\cosh[(N-1)h] y^{N+2}+ \nonumber \\
& + & \frac{2 \epsilon}{\kappa} \cosh[(N-1)h] y^N+\frac{\epsilon^2}{\kappa^2}y^2-1 \equiv \sum_{n=0}^{2N+2} a_ny^n.
\end{eqnarray}
 Note that $P(y)$ given by Eq.(A.8) belongs to the class of self-inversive (anti-palindromic) polynomials \cite{ruff1,ruff2}, i.e. one has $a_{2N+2-n}=-a_{n}$ for any $n=1,2,...,N+1$. Two roots of $P(y)$ are given by $y= \pm 1$, as can be readily check from the form of Eq.(A.8), so that $P(y)$  can be factorized as $P(y)=(y^2-1)Q(y)$, where $Q(y)$ is a self-inversive (palindromic) polynomial of order $2N$. The roots $y= \pm 1$ can not be accepted, since they would correspond to a vanishing solution ($A=B=0$) of the eigenvector. The eigenvalues $E^{\prime}$ are thus obtained from Eq.(A.3), i.e. from the relation
 \begin{equation}
 E^{\prime}= \kappa \left(y+\frac{1}{y} \right)
 \end{equation}
where $y$ is one of the $2N$ roots of the self-inversive polynomial $Q(y)$. Note that, if $y$ is a root of $Q$, then also $1/y$ is a root of $Q$, so  that according to Eq.(A.9) the number of eigenvalues $E^{\prime}$ of $H_2$ are $N$, as it should be. Note also that the energy $E^{\prime}$ is real when $|y|=1$, so that an the spectrum of $H_2$ is entirely real whenever $|y|=1$ for any root of $Q(y)$. For $\epsilon=0$, one readily obtains for $E^{\prime}=E_n$ the unperturbed values $E_n$ given by Eq.(19) in the main text. All such eigenvalues correspond to $|y|=1$, i.e. all the roots of the self-inversive polynomial $Q(y)$ are on the unit circle (they are unimodular). As $\epsilon$ is increased, the roots of $Q(y)$ remain on the unit circle, implying that the energy spectrum $E^{\prime}$ remains entirely real. However, the position of the roots $y$ on the unit circle vary as $\epsilon$ is increased from zero, indicating that the perturbation changes (mixes) all the unperturbed energies. The condition for the 
self-inversive polynomial $Q(y)$ [or equivalently $P(y)$] to have the entire roots in the unit circle, i.e. the spectrum of $H_2$ to remain real, is given by certain general theorems of number theory (see, for instance, \cite{ruff1,ruff2}). According to the theorem stated in Ref.\cite{ruff2}, an upper boundary of $\epsilon$ is obtained from the inequality 
\begin{equation}
\left( \frac{\epsilon}{\kappa} \right)^2+2 \left( \frac{\epsilon}{\kappa} \right) \cosh [(N-1)h] < 1.
\end{equation}
Note that, for a sufficiently long chain such that $\cosh[(N-1)h] \gg 1$, the first term on the left hand side of Eq.(A.10) can be neglected, so that one obtains the following limit for the perturbation strength $\epsilon$ to ensure an entry real energy spectrum of $H_2$
\begin{equation}
\epsilon < \frac{\kappa}{2 \cosh [(N-1)h]} \equiv \epsilon_c. 
\end{equation}
Numerical computation of the eigenvalues shows that, as $\epsilon$ approached the critical value $\epsilon_c$ from below, two energies $E^{\prime}$ on the real axis coalesce, and further increasing $\epsilon$ above $\epsilon_c$, the real energies bifurcate into complex conjugate energies via an EP. As $\epsilon$ is further increased, other pairs of real energies coalesce and bifurcate into complex conjugate energies, until all the energy spectrum become complex.

\end{document}